\documentclass[aps,twocolumn,pre]{revtex4-1}

\usepackage{graphicx}
\usepackage{amsmath,mathtools}
\usepackage{color}

\usepackage[hidelinks]{hyperref}

\begin{document}

\title{Learning coherences from nonequilibrium fluctuations in a quantum heat engine}

\author{Manash Jyoti Sarmah}

\author{Himangshu Prabal Goswami}
\email{mjsarmah@gauhati.ac.in,hpg@gauhati.ac.in}
\affiliation{Department of Chemistry, Gauhati University, Jalukbari, Guwahati-781014, Assam, India}
\date{\today}

\begin{abstract} 
We develop an efficient machine learning protocol to predict the noise-induced coherence from the nonequilibrium fluctuations of photon exchange statistics in a quantum heat engine. The engine is a four-level quantum system coupled to a unimodal quantum cavity.  The nonequilibrium  fluctuations correspond to the work done during the photon exchange process between the four-level system and the cavity mode. We specifically evaluate the mean, variance, skewness, and kurtosis for a range of engine parameters using a full counting statistical approach combined with a quantum master equation technique. We use these numerically evaluated cumulants as input data to successfully predict the hot bath induced coherence. A supervised machine learning technique based on K-Nearest Neighbor(KNN) is found to work better than a variety of learning models that we tested.   
\end{abstract}

\maketitle

\section{Introduction}

Non-equilibrium fluctuations in quantum systems arise when a system is driven away from thermal equilibrium, either by applying an external field or through interactions with other systems. In such situations, the system's energy and other properties fluctuate randomly over time leading to a distribution of the systems' observable values \cite{RevModPhys.81.1665,Seifert_2012,RevModPhys.83.771}. 
%In quantum systems, these fluctuations can become amplified and can have significant effects on the system's behavior   such as altering the flow of particles or energy through the system, enhancing or suppressing phase transition dynamics etc \cite{}. Inherent system properties such as coherence, correlations and entanglements also affect such fluctuations\cite{}. This can result in the decay of quantum states, leading to a loss of information and the inability to perform certain quantum computations \cite{}. 
The study of fluctuations in nonequilibrium quantum systems led to the development of quantum thermodynamics whose foundations are based on the fluctuation theorems and thermodynamic uncertainty relationships, quantified by the distributions of fluctuating quantities like heat, particle or work \cite{RevModPhys.81.1665,Seifert_2012,RevModPhys.83.771,NatPhys.16.15}. 

In general, the  cumulants of an observable  in a nonequilibrium system are studied using various theoretical and computational methods, such as the non-equilibrium Green's function formalism or the quantum master equation approach and large deviation theories \cite{RevModPhys.81.1665,Seifert_2012,RevModPhys.83.771,NatPhys.16.15,touchette2009large,wang2014nonequilibrium}. Quantitatively, the relationship between the observables  and its moments' or  cumulants is encoded in their generating functions that obey linear shift symmetries guaranteeing the validity of detailed balance \cite{wang2014nonequilibrium,Galla-Cohen,dhar2}. Using such theoretical techniques, behavior reminiscent of the dynamical first-order phase transition in absorption refrigerators has been proposed \cite{holubec2019effects} and a bold prediction on the Carnot efficiency to have the lowest probability has been established \cite{verley2014unlikely}.  The first cumulant or the nonequilibrium flux has been shown to lead to a linear
increase in steady-state coherences in spin systems \cite{PhysRevB.95.144306}. These cumulants are now experimentally measurable \cite{utsumi2010bidirectional, NewJPhys.23.065004}, eg. in cavity-mediated optomechanical resonators and even an NMR setup \cite{yang2020phonon,PhysRevA.100.042119}. Nonequilibrium dynamics have also been experimentally observed in cold Li and Na atomic systems using an optical dipole trap \cite{hegde2022non}.  

Another interesting quantity in such nonequilibrium quantum systems is the noise or bath-induced coherence resulting from Fano interference between transitions from degenerate or near-degenerate energy levels in the presence of temperature gradient or difference in noise intensities \cite{scully2011quantum,holubec2018effects}. Such coherences are not to be confused with isolated system coherence \cite{tscher}. Enhancement of flux, power, and efficiency by tuning such coherences has been theoretically predicted \cite{scully2011quantum,holubec2018effects,holubec2019effects, UHeplQHE, goswami2013thermodynamics, un, traped-ion}, esp. in quantum analogs of photocells, heat engines, and refrigerators \cite{holubec2018effects,kosloff2014quantum}.  
Observables in such devices and engines  have a nontrivial dependence on the moments and cumulants. Although standard theoretical techniques can map the effect of these coherences on the fluctuations, analytical understanding is restricted to specific physical regimes and the first moment or the flux \cite{holubec2019effects}. Higher order fluctuations, although obtainable, remain numerical and hence depend on the system specifications \cite{giri}. Some of such theoretically predicted noise-induced coherence effects have now been experimentally demonstrated, eg. in polymer solar cells, N-vacancy-based engines, pump-probe measurements \cite{brit-nat,uzdin,qutub}. However, a setup with fully controllable noise-induced coherence hasn't yet been achieved as of now.  In the current experimental quantum heat engines (QHE), two lasers mimic the thermal baths, as demonstrated in a cold Rb atomic setup \cite{Bouton-} using microwaves. Thus the parameters of the laser can be used to control the QHE specifications\cite{qutub}. Once the laser parameters are fixed  the effect of coherences on the QHE's thermodynamic quantities like flux, power, efficiency, and their fluctuations could be  explored by subsequently varying the engine specifications. Full control of all the system and bath parameters could offer intriguing new avenues for quantum control which will be relevant to other fields of the study looking at the role of coherences  in quantum processes as well as work extraction by thermal machines \cite{Zhang_2022}.

Specifying the nonequilibrium systems' specification, hence, allow us to understand and calculate the fluctuations through the moments or cumulants and gain deeper insights \cite{rudge2019counting,gong2022nonequilibrium}. However, the reverse mapping or relationship, i.e obtaining relevant information about the quantum systems' properties from the nonequilibrium fluctuations is hitherto a completely different story.   This question is highly applicable to a nonequilibrium system where the initial system or experimental setup isn't fully controllable, eg.  with Fano interferences, and is the subject behind this work. We aim to predict the noise-induced coherence from known values of the nonequilibrium fluctuations, i.e the cumulants using machine learning (ML). We choose a well-studied QHE model to demonstrate this idea \cite{PhysRevA.86.043843,scully2011quantum,dorfman}. The study of nonequilibrium quantum system properties in thermodynamics and transport using ML is a growing field of research \cite{Erdman-}. Recently reinforcement learning approach has been employed to reduce entropy production in a two-particle system \cite{sgroi}. Further, one of us showed the use of artificial neural networks to evaluate fluctuations from system parameters \cite{giri}. Also, finite time dynamics of a quantum dot engine have been assessed using Pontryagin minimum principles \cite{https://doi.org/10.48550/arxiv.2207.13104}.  

 The paper is organized as follows. In Sec. (\ref{model}), we discuss the basic structure of the QHE and the thermal bath or noise-induced coherences.  In Sec. (\ref{sec-fcs}), we numerically evaluate the cumulants using a standard full-counting statistical approach (FCS). In Sec. (\ref{model}), we show how we develop a many-to-one classification  mapping between the cumulants (nonequilibrium fluctuations) and the noise-induced coherence using ML, after which we conclude.

\section{Quantum Heat Engine: Model}
\label{model}
\begin{figure}[tbh!]
    \centering
    \includegraphics[width=7.5cm]{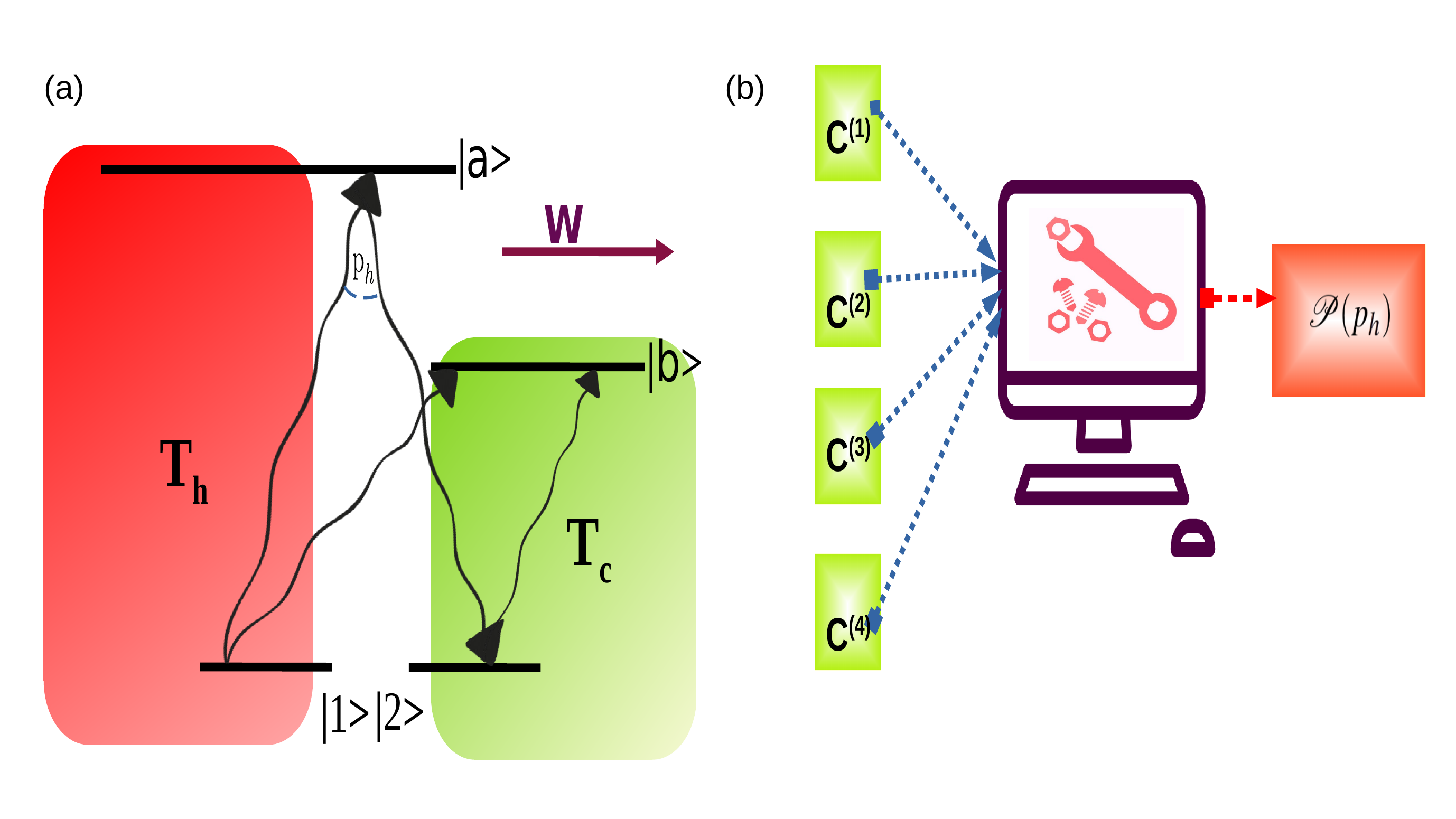}
    \caption{(a) Schematic plot of a QHE with four energy levels. Two degenerate states $|1\rangle $ and $|2\rangle $ are coupled with the higher energy states $|a\rangle $ and $|b\rangle $ through thermal baths. Hot and cold bath temperatures are labeled as $T_h$ and $T_c$ respectively. Orientation of the individual dipole is given by $p_h$. (b) Schematic plot showing the mapping between the cumulants and hot bath coherence ($p_h$). }
    \label{fig-qhe}
\end{figure}

In our QHE, two thermal baths at temperatures $T_h$ and $T_c (< T_h)$ couple asymmetrically to four quantum levels. The thermal baths can be two lasers and the quantum levels can belong to a quantum dot or molecular system \cite{microwave,un}. The upper two states couple to a unimodal cavity as shown schematically in Fig. \ref{fig-qhe}(a).
Such a type of QHE has been experimentally demonstrated in a cold atomic setup \cite{microwave,Harris2,zou}. Such QHEs can also be realized in  low-dimensional materials with high thermal conductivity \cite{zhang-2019}.
The total Hamiltonian of the four-level QHE is $\hat{H}_{T}\,=\,\hat{H}_{0}+\hat{V}_{sb}+\hat V_{sc}$, where
\begin{align}
 \hat H_0 &= \sum_{\nu\,=\,1,2,a,b}
        E_{\nu}|\nu \rangle\langle \nu |+\displaystyle\sum_{k\in h,c}\epsilon_{k}\hat{a}_{k}^{\dag}\hat{a}
        _k+\epsilon_\ell\hat{a}_{\ell}^{\dag}\hat{a}_{\ell},\\
\hat V_{sb}&=\sum_{x\,\in\,h.c}\sum_{i\,=\,1,2}\sum_{k\,=\,a,b}\Gamma_{ix}\hat{a}_{x}|k\rangle\langle i|+h.c\\
\hat V_{sc}&=g\hat{a}_{\ell}^\dag|b\rangle\langle a|+h.c.
\end{align}

 $\epsilon_{k}$, $\epsilon_{\ell}$ and $E_{\nu}$ denote the energy of the $k$-th mode of the two thermal reservoirs, the unimodal cavity and system's $\nu$-th energy level respectively. The system-reservoir coupling of the $i$-th state with the $x$-th mode of the reservoirs is denoted by
$\Gamma_{ix}$. 
The two thermal baths are assumed to be harmonic in nature and $\hat{a}^\dag (\hat{a})$ are the bosonic creation (annihilation) operators. 
The radiative decay originating from the transition $|a\rangle\to|b\rangle$ is the work done by the engine.
%%%%%%%%%%%%%%%%%%%%%%%%%%%%%%%%%
%%%%%%%%%%%%%%%%%%%%%%%%%%%%%%%%%
%%%%%%%%%%%%%%%%%%%%%%%%%%%%%%%%%
The full theoretical formalism of the equations of motion based on a Markovian master equation already exists. The full counting statistical (FCS) formalism is also already known \cite{harbola}. From these earlier works, we can express the reduced density vector in the Liouville space to be composed of four coupled populations and a quantum coherence term,  given by $|\rho\rangle = \{\rho_{11},\rho_{22},\rho_{bb},\rho_{aa},\Re(\rho_{12})\}$, with $i=1,2,b,a$. $\Re(\rho_{12})$ is the noise  or bath-induced quantum coherence between the upper two states $|1\rangle$ and $|2\rangle$.  The coherence $\Re(\rho_{12})$ between the  states $|1\rangle$ and $|2\rangle $ arise due 
to the asymmetric coupling of the system's many-body states with the hot and the cold baths. When photons are exchanged between the upper two levels and the cavity with a positive flux, there is net work done by the QHE. 

To quantify the statistics of work done in terms of photon exchange, a twisted generator can be derived which governs the time evolution of the QHE's reduced density vector and allows effective tracking of the number of photons exchanged \cite{harbola}, as per the equation $|\dot\rho(\lambda,t)\rangle = \breve{{\cal L}}(\lambda)|\rho(\lambda,t)\rangle$, where  $\lambda$ is a field that counts the number of photons exchanged between the manybody states and the cavity. 
The twisted moment generating superoperator, $\breve{\mathcal{L}} (\lambda)$ reduces to a standard Lindblad operator when $\lambda=0$. Quantitatively, $\breve{\mathcal{L}} (\lambda)=$,
\begin{small}
\begin{equation}
\label{Louv-eq}
\begin{pmatrix}
-\displaystyle\sum_{x}\Gamma_{1x}^{}n_x^{}&0&\Gamma_{1h}^{}\tilde{n}_h&\Gamma_{1c}^{}\tilde{n}_c&-2\Gamma_{12}^{}\\
0&-\displaystyle\sum_{x}\Gamma_{2x}^{}n_x^{}&\Gamma_{2h}^{}\tilde{n}_h&\Gamma_{2c}^{}\tilde{n}_c&-2\Gamma_{12}^{}\\
\Gamma_{1c}^{}n_c&\Gamma_{2c}^{}n_c& 
-\Gamma_h^{}\tilde n_h^{}-g^2\tilde n_\ell^{}&g^2 n_\ell e^{-\lambda}&\Gamma_{12c}^{}n_c\\
\Gamma_{1h}^{}n_h^{}&\Gamma_{2h}^{}n_h^{}&g^2_{}\tilde n_\ell^{}e^{\lambda}&-g^2n_\ell-\Gamma_c^{}\tilde{n}_c &2\Gamma_{12h}^{}n_h\\
-\Gamma_{12}^{}&-\Gamma_{12}^{}&\Gamma_{12h}^{}\tilde{n}_h&2\Gamma_{12c}^{}\tilde{n}_c&\bar g-\tau
\end{pmatrix}\\[2mm]
\end{equation}
\end{small}

where, $\Gamma_x= \Gamma_{1x}+\Gamma_{2x}, x =h,c.  \Gamma_{ix}=\pi\Omega|g_{ix}^{}|^2/2, i =1,2$ represent rates of transition between the states and are proportional to the modulus square of the transition dipole between the states 
$|1\rangle (|2\rangle)$ and $|a\rangle$ or $|b\rangle$, $\bar g = -n_h(\Gamma_{1h}+\Gamma_{2h})/2-
 n_c(\Gamma_{1c}+\Gamma_{2c})/2$ and $\Gamma_{12}=(\Gamma_{12c}n_c+\Gamma_{12h}n_h)/2$ with $\tau$ being a phenomenologically introduced dimensionless pure dephasing parameter \cite{scully-,UHeplQHE,himang}. $n_{c(h)}=1/\{\exp(\beta_{c(h)}(E_{b(a)}-E_1))-1\}$ are the Bose-Einstein distributions with $\tilde n_x =1+n_x$.

 The mixed term $\Gamma_{12x}$ is a measure of the strength of the noise-induced coherence stored in the actual coherence term $\Re(\rho_{12})$,
\begin{eqnarray}
 \Gamma_{12x}^{}&=&\frac{\pi\Omega}{2}|\Gamma_{1x}\Gamma_{2x}|^2,~~~~x\in h,c,\\
 &=&\sqrt{\Gamma_{1x}\Gamma_{2x}|\cos\theta_x|}
 \label{eq-theta}
\end{eqnarray}
with $\theta_x$ being the angle of relative 
orientation involving the x-th bath induced transition from states $|1\rangle$ and $|2\rangle$ to  intermediate state $|a\rangle$ or $|b\rangle$. For example, in one such  bath-induced transition to state $|a\rangle$, there are two dipole vectors, one from $|1\rangle$ and one from $|2\rangle$. These two pathways interfere and their strength depends on the angle at which the two dipoles point towards $|a\rangle$. All such contributing pathways affect the populations and are contained in the term $\Re(\rho_{12})$ through the mixed coupling terms $\Gamma_{ik}$. \cite{scully2011quantum,goswami2013thermodynamics} Assuming symmetric coupling, $\Gamma_{1x}=\Gamma_{2x} = r$, Eq. (\ref{eq-theta}) can be mathematically parameterized as 
\begin{align}
\label{eq-px}
 \Gamma_{12x}&=rp_x
\end{align}
with $p_x=\sqrt{|\cos\theta_x|}$. The two dimensionless parameters $p_h$ and $p_c$  can now be regarded as a measure of the strength of either the hot or cold bath-induced coherence, 
with $0\leq p_h, p_c\leq 1$. A perpendicular orientation of the individual dipole vectors kills the coherence, i.e, $p_c=p_h=0$. A parallel orientation generates the maximum value of the coupled term, $\Gamma_{12x}$ and decouples the population from the coherences \cite{scully2011quantum,UHeplQHE, PhysRevA.86.043843}. 
However, the extent of these coherence parameters in affecting any observables of the engine needs explicit evaluation of the concerned observable since the dependence of observables on the coherence parameters is not straightforward. Eg. the flux and the output power into the engine's unimodal cavity have been analytically shown to be nonlinearly dependent on both $p_c$ and $p_h$   within a perturbative framework \cite{scully2011quantum,UHeplQHE,goswami2013thermodynamics}. For an observed value of output power evaluated from fixed engine parameters, these noise-induced coherences  can be estimated and the relative orientations can be known. It requires knowledge of the complete QHE specifications including details of the two thermal baths, which in most of the experiments are external lasers \cite{PhysRevLett.119.050602,harris2016electromagnetically,zou2017quantum} and the relative orientation of the lasers in inducing the desired coherence.  It is well known that identifying the correct bosonic distribution temperature of lasers is a complicated experiment. \cite{zou2017quantum} Further, predicting the angles of relative orientations between the two external bath (laser) induced coherence each time the experimental parameters are changed is also an expensive protocol. \cite{magana2016hanbury,henriksen2002laser}
In this work, we propose an alternative approach, where, one of the noise-induced coherence parameters ($p_h$) can be estimated from the nonequilibrium fluctuations, i.e the cumulants of photons exchanged between the four-level system and the cavity by establishing a numerical mapping between the cumulants and the coherence parameters. Cumulants  are experimentally measurable \cite{campisi2015nonequilibrium,denzler2021nonequilibrium,NewJPhys.23.065004} and one can use the values of these cumulants to predict the coherences from some already known experimental specifications. In the next section, we start by numerically evaluating the cumulants (known data that mimics experimental specifications) using the standard FCS approach.  %Note that the dependence of these  cumulants on $p_c$ and $p_h$ and cannot be evaluated analytically \cite{}. 

%%%%%%%%%%%%%%%%%%%%%%%%%%%%%%%%%
%%%%%%%%%%%%%%%%%%%%%%%%%%%%%%%%%

\section{Numerical Full Counting Statistics}

\label{sec-fcs}
The work done by the engine is the emission of photons into the cavity involving states $|a\rangle$ and $|b\rangle$. Hence, there is a net flux of photons exchanged between the system and the cavity. This exchange also results in higher-order fluctuations of photon exchange which are statistically identifiable in the form of cumulants.
In the steady state, when $\lambda= 0$, a zero eigenvalue  is obtained from the RHS of Eq.(\ref{Louv-eq}). This zero-eigenvalue  corresponds
to a cumulant generating function, $S(\lambda)$ within the domain $\lambda\in\{-\infty,\infty\}$, from which the cumulants of photon exchange can be directly evaluated as,
\begin{align}
\label{cum-eq}
 j^{(i)}&=\partial_\lambda^i S(\lambda)|_{\lambda=0}.
\end{align}
Here, $i=1,2,3$ and $4$ correspond to the mean, variance, skewness, and kurtosis respectively, and are affected by the noise-induced coherences.  Whenever $p_c=p_h=0$, the effect of coherences vanish since $\Re (\rho_{12})=0$\cite{goswami2013thermodynamics}. We refer to these values as the classical values and denote the cumulants in absence of coherence ($p_c=p_h=0$) as $ j^{(i)}_o$. We also define a dimensionless ratio,
\begin{align}
 \label{dim-eq}
 C^{(i)}:=\displaystyle\frac{ j^{(i)}}{ j^{(i)}_o}
\end{align}
which signifies the extent of change in the values  of the quantum  cumulants (when $p_x\ne 0$) in comparison to the classical case. When $|C^{(i)}|>(<)1$, coherences increase (decrease) the value of the cumulant in comparison to the classical case. Note that $C^{(1)}$ can be evaluated analytically within a perturbative regime\cite{scully2011quantum} without evaluating $S(\lambda)$. 
%%%%%%%%%%%%%%%%%%%%%%%%%%%%%%%%%%%%%
%%%%%%%%%%%%%%%%%%%%%%%%%%%%%%%%%%%%%
%%%%%%%%%%%%%%%%%%%%%%%%%%%%%%%%%%%%%
%%%%%%%%%%%%%%%%%%%%%%%%%%%%%%%%%%%%%
\begin{table}[h!]
  \begin{center}
    \caption{Range of Liouvillian parameters.  Other system parameters were fixed at $E_1=0.5$, $E_a=3.0$, $E_b=2.0$, $g=1.0$, $r=0.1$ and $\tau=0.1$. It should be noted that these parameter values were used for the QHE throughout the entire paper. Units are in $k_B \rightarrow 1$, $\hbar \rightarrow 1$ (atomic units).} 
\begin{tabular}{c|c}
\label{tab-range-data}
     % \toprule % <-- Toprule here
      \textbf{Parameters} & \textbf{Value} \\
   % \midrule % <-- Midrule here
      \hline
      $T_c$ & 0.4-2.5 \\
      $T_h$ & 3.0-4.5 \\
      $T_l$ & 1.0-7.0 \\
      $p_c$ & 0.0-1.0 \\
      $p_h$ & 0.0-1.0 \\
     % \bottomrule % <-- Bottomrule here
    \end{tabular}
  \end{center}
  
\end{table}
%%%%%%%%%%%%%%%%%%%%%%%%%%%%%%%%%%%%%
%%%%%%%%%%%%%%%%%%%%%%%%%%%%%%%%%%%%%
%%%%%%%%%%%%%%%%%%%%%%%%%%%%%%%%%%%%%

Each of these cumulants is affected by the coherences. From a perturbation theory standoff, the dependence of the flux on these coherences is well understood since  analytical relationships between the two exist \cite{goswami2013thermodynamics}. Optimization of flux via coherences is a relatively accepted and well-understood phenomenon \cite{scully2011quantum}. However, the higher-order cumulants are not obtainable analytically since a closed form of $S(\lambda)$ cannot be evaluated. Hence, each time any engine parameter is changed, the effect of coherences on these cumulants cannot be predicted  and one needs to resort to numerics to evaluate the cumulants \cite{PhysRevE.99.022104}. 

%%%%%%%%%%%%%%%%%%%%%%%%%%%%%%%%%%%%%

\section{Initializing the learning of coherence}
\label{mapp}
The central goal of this work is to use the knowledge of cumulants obtained from Eq.(\ref{dim-eq}) to predict one of the coherences, the second term in the r.h.s of Eq.(\ref{eq-px}). Therefore, from an ML point of view, the input data are the set of cumulants, $\{C^{(i)}\}$, and the output is $p_{h}^{}$. The data is numerically generated (but can be experimentally obtained), for different combinations of the QHE's parameters as indicated in Table(\ref{tab-range-data}). Usually, in an experimental set-up, the bare system is fixed, eg. a cold atomic set-up \cite{zhang2019microwave}. So, we first fix the energies of the four levels and the system-bath coupling values. Since the baths are bosonic in nature, the effective bath temperatures are controllable and we choose to vary these for the entire range of $p_c$ and $p_h$ values, indicated in Table \ref{tab-range-data}. Note that, there is no restriction on the parameters to be varied and one can choose to include all the system parameters to evaluate the cumulants.  We first attempted to predict the values of the hot bath-induced coherence parameter, $p_h$, from the cumulants by numerically identifying the unknown many-to-one mapping.  As  a first guess, we can define a regression, $f_n:\{C^{(i)}\}\rightarrow p_h$, where $i\in \{1,2,3,4\}$ is the cumulant index with $f_n$ being the desired many to one mapping function. Interestingly, standard ML techniques were unable to learn the mapping from the numerically calculated cumulant values (See supplementary information).  This can be bypassed if we treat the mapping as a classification problem by dividing the entire range of $p_h$ values into several mathematical intervals. Supervised forecasting of the pre-defined intervals of hot bath-induced coherence parameter $p_h$ from the set of the cumulants is what we proceed to do.
\subsection{Classification Mapping}
In our work, we choose four such intervals (which are usually referred to as classes and are indicated in Table(\ref{tab-class})). This approach allows us to define a multi-class classifier mapping between the cumulant space and the noise-induced hot-bath coherence intervals or classes. Mathematically, we define the classifier as follows:

\begin{align}
\label{eq-map}
 f_c:\{C^{(i)}\}\rightarrow \mathcal P(p_h),
 \end{align}
 where $f_c$ is the multi-class classifier that maps the cumulants, represented by the set $\{C^{(i)}\}$, to the four different sets of pre-defined intervals of the hot bath-induced coherence values. $\mathcal{P} (p_h)$ is the probability of observing each of the four classes of $p_h$ values whose ranges are indicated in Table(\ref{tab-class}).

\begin{table}[h!]
  \begin{center}
    \caption{Four intervals of the hot-bath induced coherence parameter $p_h$.} 
\begin{tabular}{c|c}
\label{tab-class}
     % \toprule % <-- Toprule here
      \textbf{$p_h$ range} & \textbf{Class} \\
   % \midrule % <-- Midrule here
      \hline
      0.00-0.25 & Class 0 \\
      0.25-0.50 & Class 1 \\
      0.50-0.75 & Class 2 \\
      0.75-1.00 & Class 3 \\
     % \bottomrule % <-- Bottomrule here
    \end{tabular}
  \end{center}
  
\end{table}

\begin{figure}[bth!]
    \includegraphics[width=7.5cm]{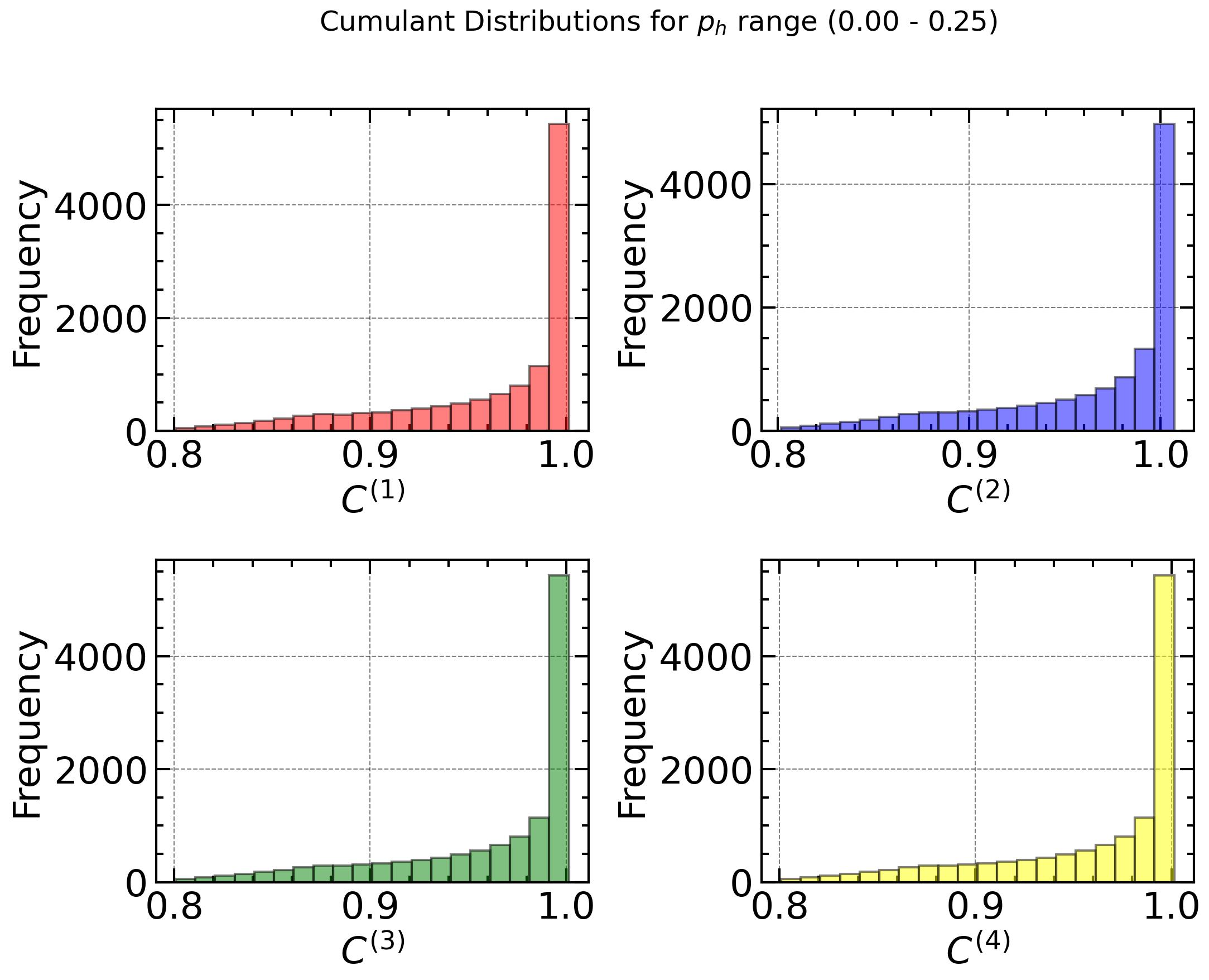}
    \caption{Histogram of cumulants for class 0. The x-axis represents the different cumulants and the y-axis represents the frequency of occurrence of the cumulants in each class evaluated using Eq.(\ref{dim-eq}) for parameters in Table (\ref{tab-range-data}). }
    \label{0dist}
\end{figure}

\begin{figure}[bth!]
    \includegraphics[width=7.5cm]{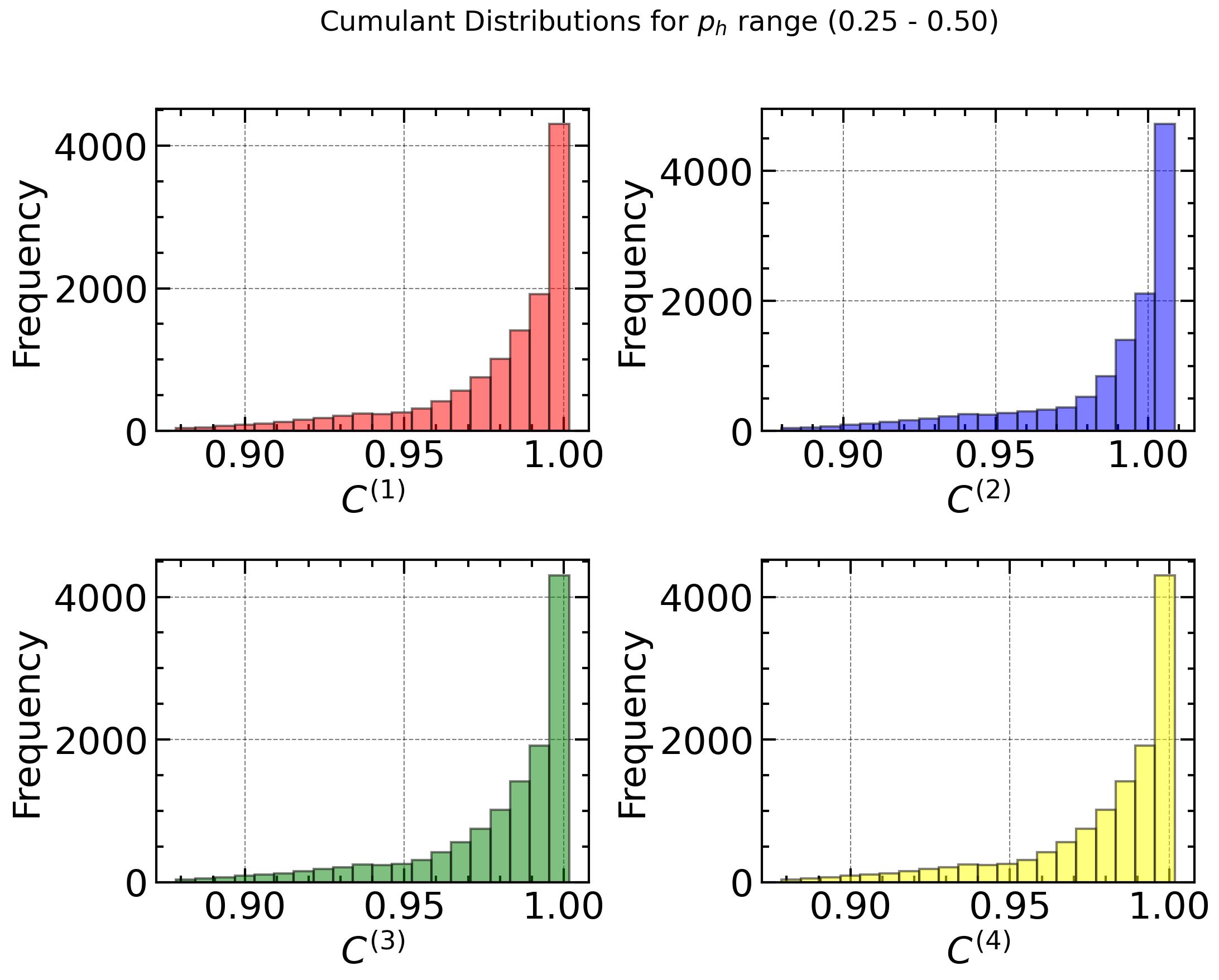}
    \caption{Histogram of cumulants for class 1. The x-axis represents the different cumulants and the y-axis represents the frequency of occurrence of the cumulants in each class evaluated using Eq.(\ref{dim-eq}) for parameters in Table (\ref{tab-range-data}).  }
    \label{1dist}
\end{figure}

\begin{figure}[bth!]
    \includegraphics[width=7.5cm]{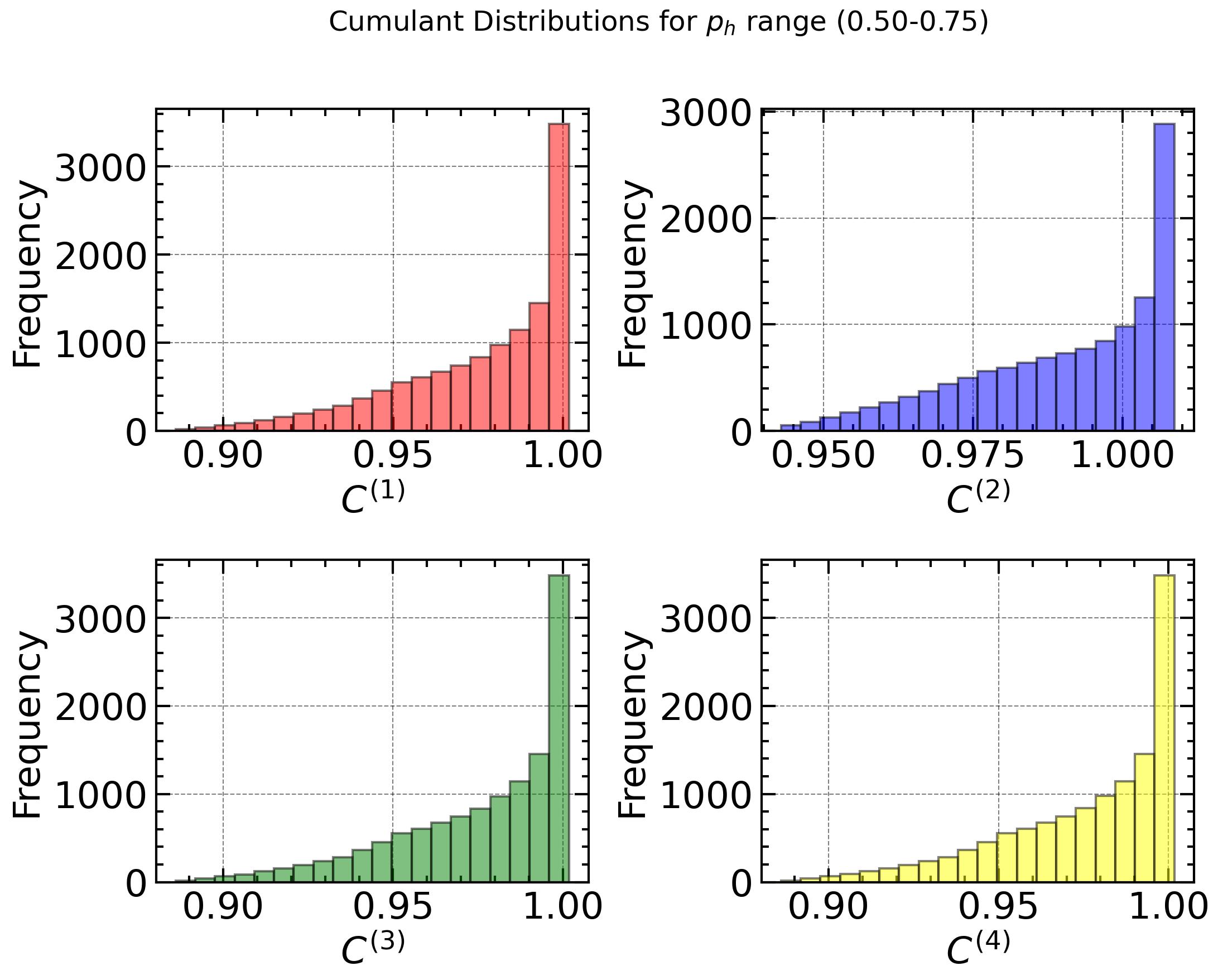}
    \caption{Histogram of cumulants for class 2. The x-axis represents the different cumulants and the y-axis represents the frequency of occurrence of the cumulants in each class evaluated using Eq.(\ref{dim-eq}) for parameters in Table (\ref{tab-range-data}).  }
    \label{2dist}
\end{figure}

\begin{figure}
    \includegraphics[width=7.5cm]{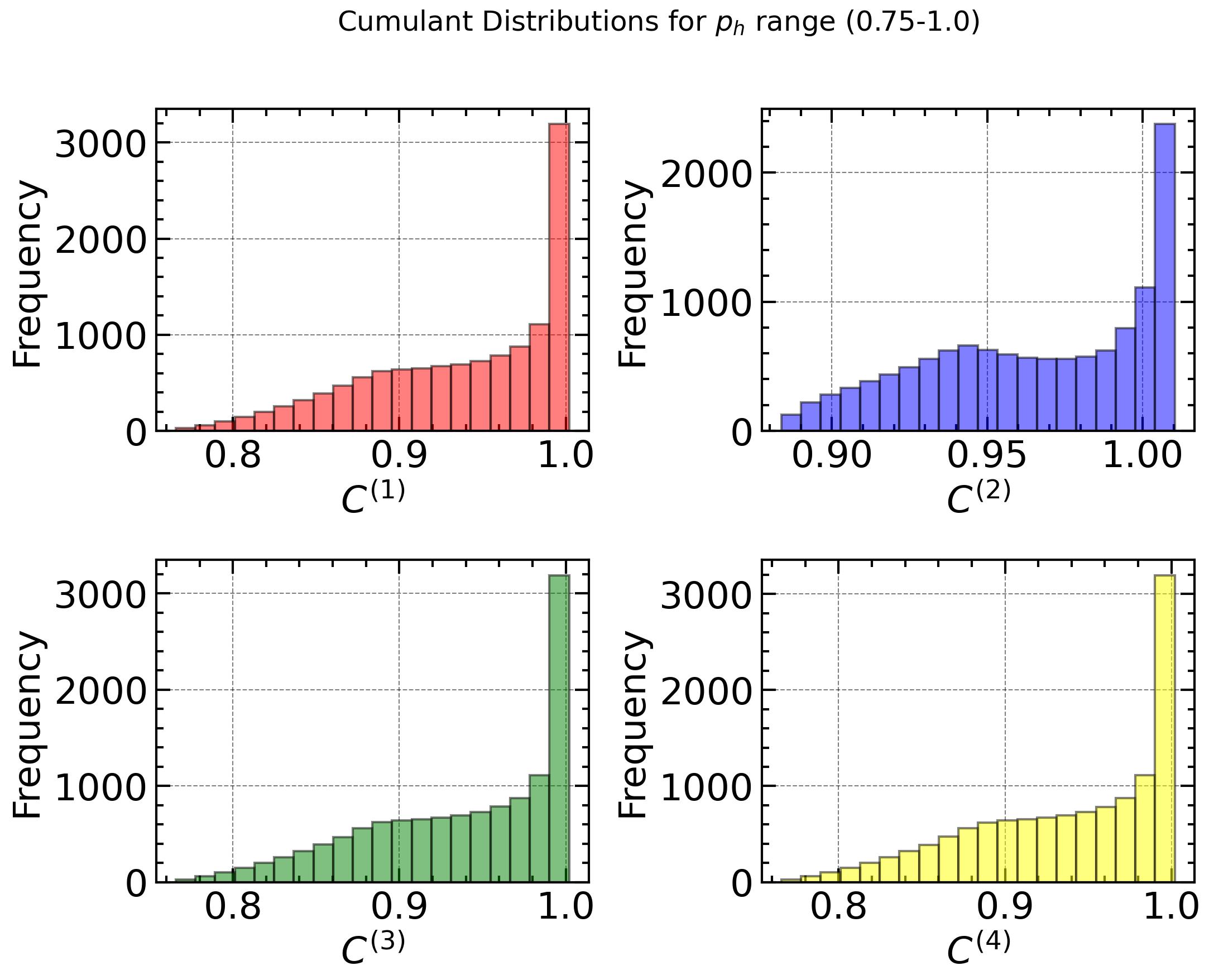}
    \caption{Histogram of cumulants for class 3. The x-axis represents the different cumulants and the y-axis represents the frequency of occurrence of the cumulants in each class evaluated using Eq.(\ref{dim-eq}) for parameters in Table (\ref{tab-range-data}).  }
    \label{3dist}
\end{figure}

We numerically evaluate the first four cumulants using Eq.(\ref{dim-eq}) and parameters from Table (\ref{tab-range-data}). We manually distribute each of the evaluated dimensionless cumulants to their respective $p_h$ classes to initialize our training process. We use this data for training our ML models. The distribution of each of these four cumulants in the four $p_h$ classes is shown in Figure \ref{0dist}-\ref{3dist}.


In the next section, we begin by performing a preliminary investigation of the performance of several supervised ML models by including all four sets of cumulants. Each of the chosen ML models will aim to establish the following relation,
\begin{align}
\label{f1-map}
     f_1:\{C^{(1)}, C^{(2)}, C^{(3)}, C^{(4)}\}\rightarrow \mathcal{P} (p_h).
\end{align}

We initially choose eight popular ML algorithms —  Logistic Regression(LR), K-Nearest Neighbour Classifier(KNN), Decision Tree Classifier (DT), Support Vector Classifier(SV), Random Forest Classifier(RF), Ada Boost Classifier(AB), Gradient Boosting Classifier(GBC) and Gradient Boosting Naive Bayes(GBN)\cite{GBR,Lr,svr,mlpr} to do a preliminary comparison on the performance of each of the models on different sizes of the generated data. 
The data is supplied to the ML algorithm so as to initiate the learning phase using Python's scikit-learn module \cite{scikit-learn}.

\subsection{ML Model Performance} 
\label{sec-ml-dev}
\begin{figure}[tbh!]
    \includegraphics[width=7.0cm]{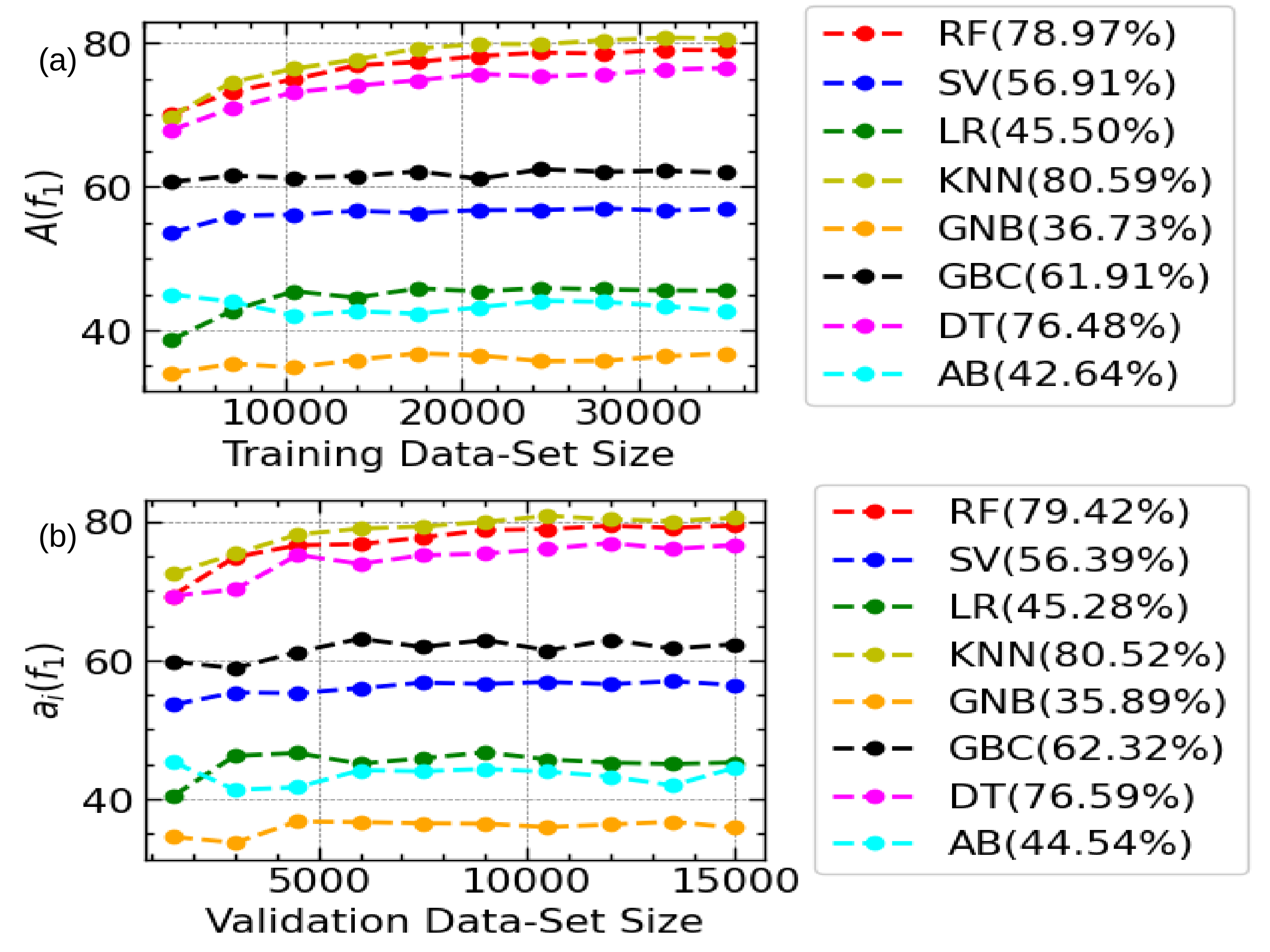}

    \caption {(a)A($f_1$)of the trained models with varying data-set size after 5-fold validation. (b) $a_{i}(f_1$) of the validated models. The saturation values of the models are indicated in the brackets.}
    \label{rmse-graph}
\end{figure}

To assess the performance of the eight models' ability to make predictions, we look at a model performance indicator, $a_i(f_c)$ (accuracy of a particular classifier, $f_c$ ), which is the accuracy percentage in a single-shot trial,
\begin{align}
\label{eq-accuracy}
    a_i(f_c) &=\frac{1}{N}\sum_{j=1}^{N}\delta_{y_{j}\hat y_{j}}\times 100\% .
\end{align}
Here, N is the total number of samples in the training/validation set, $y_j$($\hat{y}_j$) is the actual (predicted) class for the $j$-th numerically evaluated cumulant set. $\delta_{y_{j}\hat y_{j}}$ is the kronecker delta. To maximize the single-shot accuracy, we used  a 5-fold cross-validation technique\cite{k-fold-cv}. The average performance across all 5 iterations is used as the model's overall performance. The accuracy of the model can be represented as the mean of $a_{i}(f_c)$ obtained in each iteration, which describes the overall accuracy as:
\begin{align}
   A(f_c)= \frac{1}{5}\sum_{i=1}^{5} a_i(f_c).
\end{align}
We used scikit-learn library's KFold() class to create the 5-fold cross-validation iterator and use it in combination with the cross\_val\_score() function to evaluate each model's accuracy\cite{scikit-learn} after 5-fold validation for Eq.(\ref{f1-map}).\\
We first evaluate $A(f_1)$ of the four-to-one model, Eq.(\ref{f1-map}), for each of the eight ML algorithms for different training and validation data with a split ratio of 70:30 (training: validation). The establishment of the
models and their performance is evaluated using
Python's scikit-learn module. The hyperparameters of all ML models are initially set to default values.
When the total sample size reaches 50,000 (i.e. 35,000 training data and 15,000 validation) the $A(f_1)$ value saturates. The $A(f_1)$ values for the training and validation sets are shown in Figure (\ref{rmse-graph}(a) and (b) respectively). 

The KNN and the RF model are found to have the highest value of $A(f_1)$, for the training (validation) set saturating at 80.59(80.52) and 78.97(79.42) respectively. So, one may choose either KNN or RF classifier for further analysis.
To further improve the accuracy of the models we perform a hyperparameter optimization as discussed in the next section.

\subsection{Hyperparameter tuning}
KNN and RF models have several hyperparameters \cite{hyper-all,hyper-knn,hyper-rf}. As a standard procedure, three for KNN and six hyperparameters for the RF model are usually optimized prior to learning. The optimization of the hyperparameters is again done using the scikit-learn library in Python \cite{scikit-learn,research-rscv}. The RandomizedSearchCV function from the sklearn.model\_selection module was employed to perform the hyperparameter optimization. We explored a wide range of possible values for the hyperparameters as indicated in Table \ref{table:hyper}. The optimized hyperparameters were obtained using the best\_params\_ attribute of the RandomizedSearchCV object. On using the optimized hyperparameters the training (validation) $A(f_1)$ ($a_i(f_1)$) value for KNN algorithm is 81.61\%(82.04\%) and for the RF algorithm, it is 80.21\%(80.51\%). Note that, we optimized the hyperparameters for all the models and their performances are shown in the supplementary text.
We now chose the  KNN model because of its lower training and prediction times to investigate the following three-to and two-to-one multi-class classifiers,
\begin{align}
    \label{3-rel}
    f_2&: \{C^{(1)}, C^{(2)}, C^{(3)}\} \rightarrow \mathcal{P}(p_h)\\
    % \label{234-rel}
    % f_3&: \{C^{(2)}, C^{(3)}, C^{(4)}\} \rightarrow \mathcal{P}(p_h)\\
    \label{12-rel}
    f_3&: \{C^{(1)}, C^{(2)}\} \rightarrow \mathcal{P}(p_h)
    % \label{23-rel}
    % f_5&: \{ C^{(2)}, C^{(3)}\} \rightarrow \mathcal{P}(p_h).
\end{align}
 To optimize the hyperparameters of the KNN algorithm based $f_2$ and $f_3$ classifier models, we again employ the Random Search Cross Validation (RSCV) method\cite{hyperparam}. We randomly chose combinations of the three KNN hyperparameters and used the optimized values to separately train both the  classifier models $f_c, c = 2$ and $3$. The final models were chosen based on the set of hyperparameters that achieved the highest $a_{i}(f_c)$ value on the validation set as indicated in Table (\ref{table:map}). From Table(\ref{table:map}, \ref{tab:mapping-performance}) it is seen that all the classifiers achieved $a_i(f_c)$ of greater than 81\%. Using the classifier models, we now proceed to identify which among the three mappings, Eq.(\ref{f1-map},\ref{3-rel} and \ref{12-rel}) is best suited to predict the hot bath-induced coherence intervals from the nonequilibrium fluctuations.

\begin{widetext}
%\begin{widetext}
    \begin{table*}[tbh!]
    % \begin{sidewaystable}[2]
    \centering
    % \footnotesize
    \begin{tabular}{|c|c|c|c|c|c|}
    \hline
    \textbf{Model} &    \textbf{Hyperparameters} &     \textbf{Range} & \textbf{Total no.} &   \textbf{Optimised Value} & \textbf{Time}  \\ \hline
    KNN & n\_neighbours & 1 to 50 & 50 & 15     & Training/Prediction \\ \hline
     & weights & uniform and    distance\cite{hyper-knn} & 2 & distance    &  20ms$\pm$57.5$\mu$s/71.5ms$\pm$276$\mu$s  \\ \hline
    & metric & euclidean and   manhattan\cite{hyper-knn} & 2 &   euclidean &  \\ \hline
    RF & n\_estimators & 200-2000 & 10 &    1800 & 1min    56s$\pm$105ms/4.16s$\pm$5.6ms \\ \hline
    & max\_features & auto and     sqrt\cite{hyper-rf} & 2 & auto & \\     \hline
    & max\_depth & 10-110 & 11 & 90 & \\   \hline
    & min\_sample\_split & 2,5,10 & 3 & 2 &    \\ \hline
    & min\_sample\_leaf & 1,2,4 & 3 & 4 &  \\ \hline
    \end{tabular}
    \caption{Hyperparameter tuning for KNN  and RF models for $f_1$ classifier.}
    \label{table:hyper}
    % \end{sidewaystable}
    \end{table*}
% \end{widetext}

% \begin{widetext}
\begin{table}[b!]
\centering
\begin{tabular}{|c|c|c|c|c|c|c|c|}
\hline
\textbf{Mappings} & \textbf{n\_neighbors} & \textbf{weights} & \textbf{metric} & \textbf{Training Time } & \textbf{Prediction Time} & \textbf{$A_t(f_c)$} & \textbf{validation $a_{i}(f_c)$}\\ \hline
$f_1$ & 15 & distance & euclidean & 20ms$\pm$57.5$\mu$s & 71.5ms$\pm$276$\mu$s & 81.46\% & 82.16\% \\ \hline
$f_2$ & 9 & uniform & euclidean & 17.5ms$\pm$43.1$\mu$s & 253ms$\pm$4.87ms & 82.22\% & 82.11\% \\ \hline
$f_3$ & 24 & distance & euclidean & 14.8ms$\pm$65.8$\mu$s & 77.9ms$\pm$130$\mu$s & 81.64\% & 82.82\% \\ \hline

\end{tabular} 
\caption{Optimised hyperparameter for different mappings.}
\label{table:map}
\end{table}
\end{widetext}

\subsection{KNN model performance}
\begin{figure}
    \centering
   \includegraphics[width=2.5in,height=2.5in]{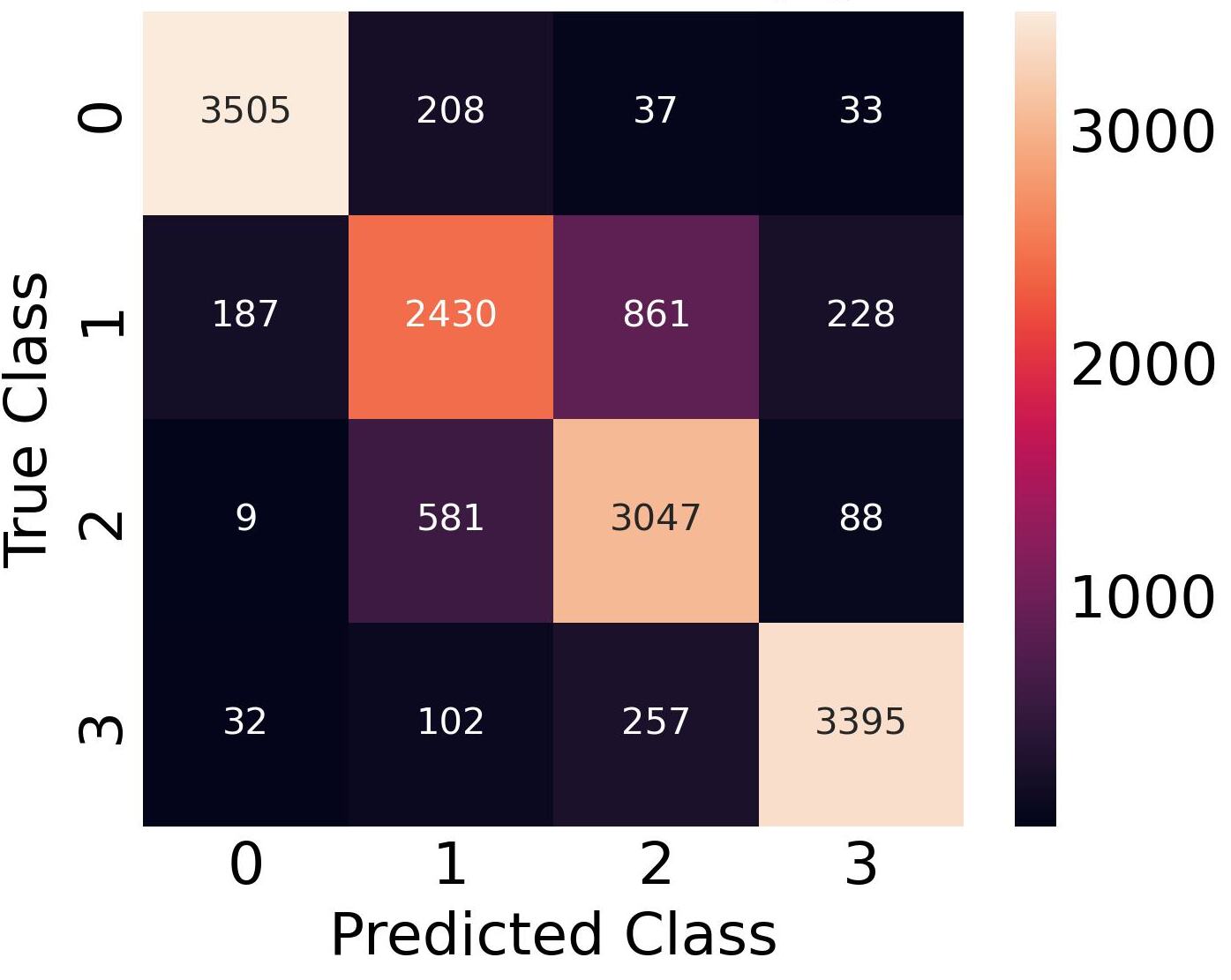}
    \caption{Confusion matrix for a 4-class classification model for mapping $f_3$. The rows represent the predicted classes and the columns represent the true classes. The entries in the matrix are the number of instances in each class. The diagonal entries represent the number of correct classifications and the off-diagonal entries represent the number of misclassifications.}
    \label{f1-img}
\end{figure}
\begin{figure}[tbh!]
    \centering
    \includegraphics[width=3.9in,height=3.0in]{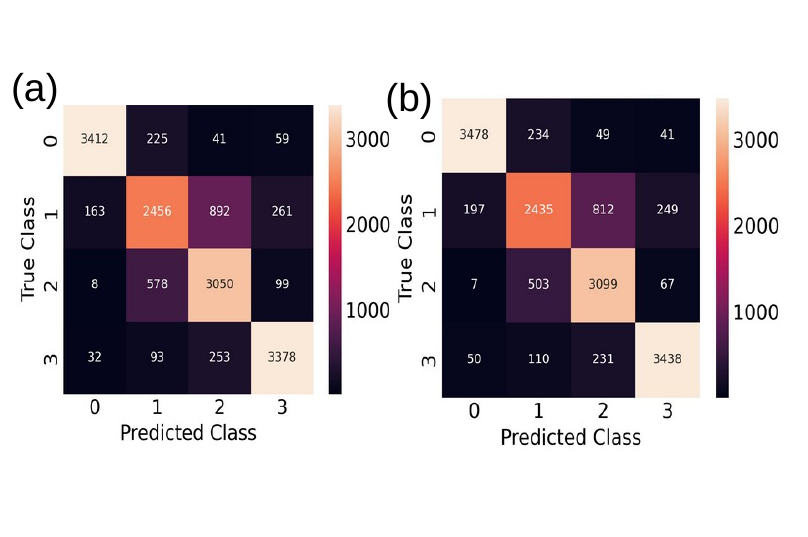}
    \caption{Confusion matrices for the mappings (a) $f_1$ and (b) $f_2$ respectively.}
    \label{confusmat}
\end{figure}

The trained KNN classifiers' (with the optimized hyperparameters) are applied to 15,000 instances(data points) of the validation data-set and their overall performance is systematically displayed in Fig \ref{f1-img}-\ref{confusmat} using a confusion matrix \cite{confusion-mat}, from which we evaluate some common performance indicators like the overall accuracy, precision, recall, F score, and Matthews correlation coefficient\cite{mcc} for all the classifiers. Denoting the confusion matrix elements for a particular classifier as $\chi_{mn}$, the overall validation accuracy (previously defined in Eq.(\ref{eq-accuracy})) can also be recast as:
\begin{align}
     a_{i}(f_c) = & \displaystyle\frac{\displaystyle\sum_{m=0}^3 \chi_{mm}}{\displaystyle\sum_{m,n=0}^3  \chi_{mn}}\times 100\%.
\end{align}
The precision($p_{k}$) and recall($R_{k}$) for the $k$-th class of a particular classifier can be expressed as:
\begin{align}
    p_{k}^{}=& \displaystyle\frac{\chi_{kk}}{\displaystyle\sum_{m=0}^3 \chi_{mk}},~~
    R_{k}= \displaystyle\frac{\chi_{kk}}{\displaystyle\sum_{m=0}^3 \chi_{km}},
\end{align}
which are now used to evaluate the F score for the $k$-th class.
\begin{align}
    F_{k}=& 2 \times \frac{p_{k} \times R_{k}}{p_{k} + R_{k}}\times 100\%
\end{align}
We also define the Matthews correlation coefficient($\phi_{k}$ ) of the $k$-th class as,
\begin{align}
    \displaystyle\phi_{k} = & \frac{(t^{+}_{k} \times t^{-}_{k} - f^{+}_{k} \times f^{-}_{k})}{\sqrt{(t^{+}_{k} + f^{+}_{k})(t^{+}_{k} + f^{-}_{k})(t^{-}_{k} + f^{+}_{k})(t^{-}_{k} + f^{-}_{k})}} \times 100\%
\end{align}
where, for the $k$-th class, $t^{+}_{k}=\chi_{kk}$ is the true positive, $f^{+}_{k}=\sum_{m\ne k}\chi_{mk}$ is the false positive and $f^{-}_{k}=\sum_{m\ne k}\chi_{km}$ is the false negative. The true negative, $t^{-}_{k}$, is equal to the sum of all elements of the co-factor matrix for the $k$-th class. For each of the classifiers defined in Eq.(\ref{f1-map},\ref{3-rel} and \ref{12-rel}), 
the performance indicators $F_{k}$ and $\phi_{k}$ are tabulated in Table(\ref{tab:mapping-performance}). Based on the results from Table(\ref{tab:mapping-performance}), we observe that predictions for the $0$-th class are the best among all the four classes, followed by class 3. We can infer that lower($p_h<0.25$) and higher values($p_h>0.75$) of $p_h$ can be predicted from the fluctuations with greater confidence than the intermediate $p_h$ values($0.25\le p_h \ge 0.75 $) with the $f_3$ classifier being the most efficient. The $f_3$ classifier also takes lesser time to train and predict in comparison to the other four classifiers and also has a higher value of overall validation accuracy ($a_i(f_3)=82.82\%$) as seen from Table(\ref{table:map}).

% \begin{widetext}
\begin{table}[tbh!]
\centering
\caption{$F$ score of the trained classifiers for each class. The Matthews Correlation Coefficients for each class are indicated in the brackets.}
\label{tab:mapping-performance}
\begin{tabular}{|c|c|c|c|c|c|}
\cline{1-4}
\textbf{}  & \textbf{$f_1$}  & \textbf{$f_2$}  & \textbf{$f_3$}    \\  \hline
\textbf{$F_0(\phi_0)$} & 92.82(90.51) & 92.33(89.76) & 93.27(91.02) \\ \hline
\textbf{$F_1(\phi_1)$} & 68.95(59.50) & 69.82(60.90) & 69.16(59.92) \\ \hline
\textbf{$F_2(\phi_2)$} & 76.53(68.33) & 78.78(71.57) & 76.88(68.84) \\\hline
\textbf{$F_3(\phi_3)$} & 89.45(85.90) & 90.19(86.85) & 90.17(86.88)\\ \hline
\end{tabular}
\end{table}
% \end{widetext}
\subsection{Application of the trained model} 
\label{application}

\begin{figure}[tbh!]
    \centering
    \includegraphics[width=0.5\textwidth,height=4.0in]{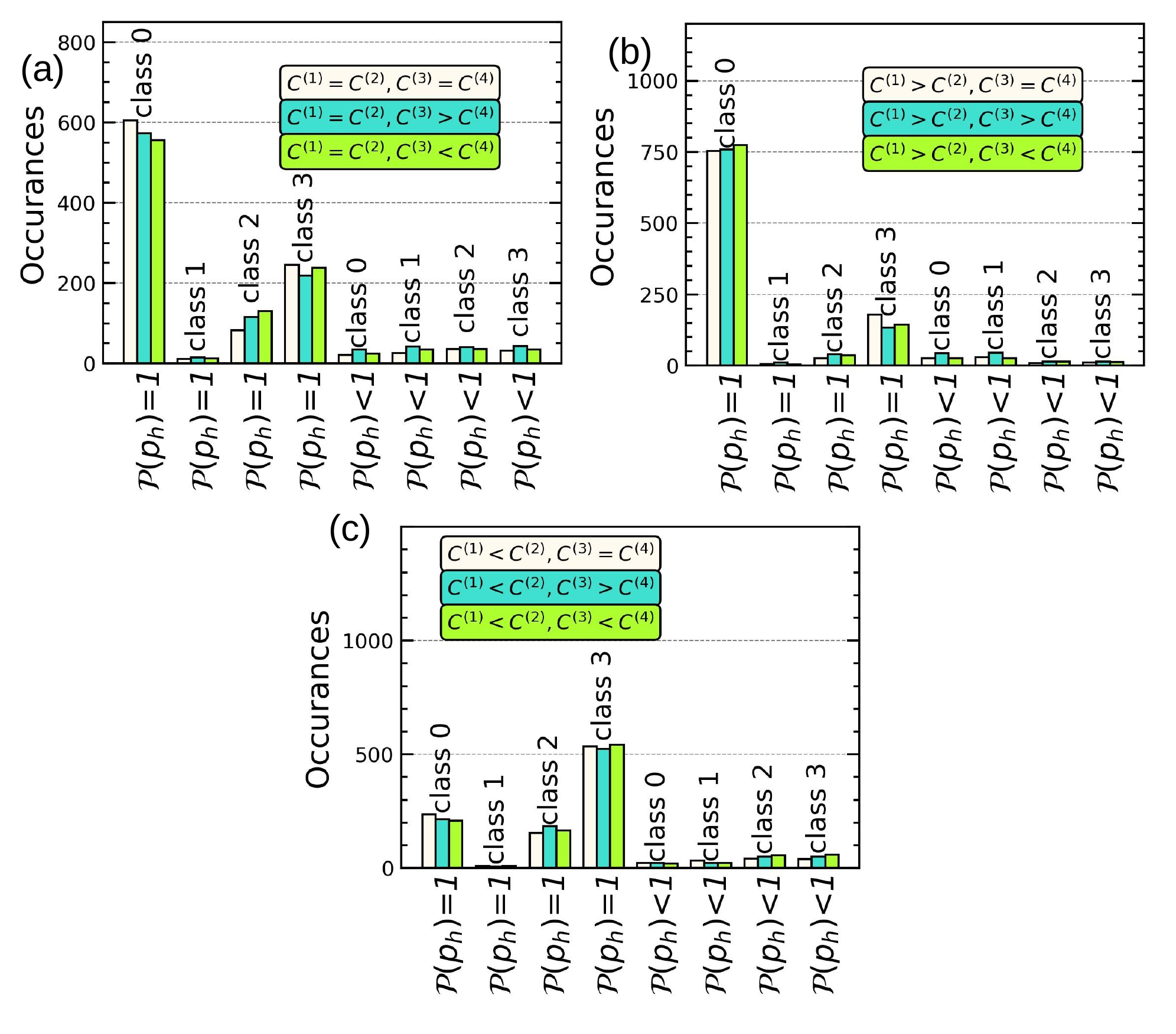}
    \caption{\textbf{$f_1$ classifier test results:} (a) When $C^{(1)}=C^{(2)}$ and $C^{(3)}$, $C^{(4)}$ is varied as shown in the inset, (b) When $C^{(1)}>C^{(2)}$ and $C^{(3)}$, $C^{(4)}$ is varied as shown in the inset,(c) When $C^{(1)}<C^{(2)}$ and $C^{(3)}$, $C^{(4)}$ is varied as shown in the inset. }
    \label{f1-test}
\end{figure}

To apply the classifiers, we generate $1000$ values for each of the input parameters $C^{(1)}$, $C^{(2)}$, $C^{(3)}$, and $C^{(4)}$ in the range of 0.76-1.001, 0.80-1.01, 0.76-1.002, and 0.76-1.001, respectively. These ranges correspond to the initial data distribution shown in Figures(\ref{0dist}-\ref{3dist}).  We use the trained classifiers to predict the probability of observing hot-bath-induced coherence in one of the four classes using the predict\_proba() function from the scikit-learn library\cite{scikit-learn}. For the $f_1$ classifier, we examined nine cases in which we impose constraints on the cumulants $C^{(1)}$, $C^{(2)}$ and $C^{(3)}$, $C^{(4)}$, as depicted in Figure \ref{f1-test}(a), (b), and (c). In Figure(\ref{f1-test}(a)), $C^{(1)}=C^{(2)}$ is kept fixed while three different constraints are imposed on $C^{(3)}$ and $C^{(4)}$ (=, >, <)as shown in the inset. The number of cases predicted by the KNN model where the probability of $p_h$ belonging to the lowest coherence interval (class 0) is greater in comparison to the other classes. This number does not vary significantly regardless of the constraints imposed on $C^{(3)}$ and $C^{(4)}$ as seen from the extreme left-hand bar plots in Figure(\ref{f1-test}(a)). The model predicted 605, 573, and 556 instances out of 1000 to be in class 0 with unit probability for the three constraints on  $C^{(3)}$ and $C^{(4)}$ respectively. Similarly, for cases where $C^{(1)}>C^{(2)}$ (Figure (\ref{f1-test}(b))), irrespective of the conditions on $C^{(3)}$ and $C^{(4)}$, the number of cases where the probability of $p_h$ lying in class 0 was higher in comparison to the rest. In Figure (\ref{f1-test}(b)), the model predicted 753, 758, and 775 instances out of 1000 to be in class 0 with unit probability. In Figure (\ref{f1-test}(c))), we impose the same set of constraints on $C^{(3)}$ and $C^{(4)}$ while maintaining $C^{(1)}<C^{(2)}$. This time the number of cases predicted by the model with unit probability was highest for class 3. The model predicted 536, 524, and 542 instances of class 3 out of 1000 instances for the three constraints respectively. Under this scenario too, the constraints put on the values of $C^{(3)}$ and $C^{(4)}$ were insignificant. 

\begin{figure}[h]
    \centering
    \includegraphics[width=5cm,height=10cm]{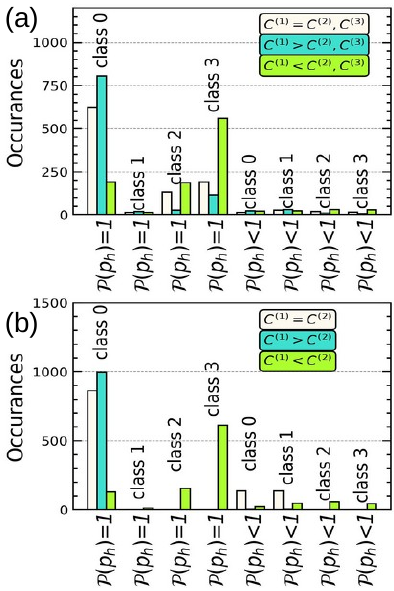}
    \caption{(a)$f_2$ classifier test result, $C^{(1)}$ and $C^{(2)}$ is varied as shown in the inset, (b)$f_3$ classifier test result, $C^{(1)}$ and $C^{(2)}$ is varied as shown in the inset.}
    \label{f2-f3-test}
\end{figure}

For the $f_2$ classifier, we exclude the $C^{(4)}$ values from the test set and investigate three cases by imposing constraints on $C^{(1)}$ and $C^{(2)}$, while also considering $C^{(3)}$ values as depicted in Figure(\ref{f2-f3-test}(a)). Here also, our findings indicate that regardless of the $C^{(3)}$ values, the number of cases predicted by the model with unit probability was highest for class 0 when $C^{(1)}=C^{(2)}$ and $C^{(1)}>C^{(2)}$. The model predicted 622 and 804 instances belonging to class 0 out of 1000 instances for the two cases respectively. However, when $C^{(1)}<C^{(2)}$, the number of cases predicted by the model with unit probability was highest for class 3. The model predicted 560 instances belonging to class 3 out of 1000 instances. For the $f_3$ classifier, we exclude the $C^{(3)}$ and $C^{(4)}$ values from the test set and investigate three cases by imposing constraints on $C^{(1)}$ and $C^{(2)}$ values as depicted in Figure(\ref{f2-f3-test}(b)). When $C^{(1)}=C^{(2)}$ and $C^{(1)}>C^{(2)}$, the number of cases predicted by the model with unit probability was highest for class 0. The model predicted 862 and 995 instances belonging to class 0 out of 1000 instances for the two cases respectively. When $C^{(1)}<C^{(2)}$, however, the number of cases predicted by the model with unit probability was highest for class 3. The model predicted 613 instances belonging to class 3 out of 1000 instances. 
The results from all the classifiers are consistent with one another. From the above results, it can be concluded that regardless of the condition on $C^{(3)}$ and $C^{(4)}$, for $C^{(1)}=C^{(2)}$ (Poissonian statistics \cite{poissonian})and $C^{(1)}>C^{(2)}$ (bunched statistics \cite{poissonian}), the number of cases predicted by all the models with unit probability was highest for class 0, i.e, there is a preference for low coherence values whenever the flux and the variance are equal or when the flux is greater than the variance, and for 
$C^{(1)}<C^{(2)}$ (antibunched statistics \cite{poissonian}), the number of cases predicted by the models with unit probability was highest for class 3, i.e, whenever the variance is greater than flux we see a preference for higher coherence values. 

\section{Conclusions}
In conclusion, we have used the k-nearest neighbors (KNN) algorithm to classify the hot bath-induced quantum coherence values based on the cumulants of a four-level quantum heat engine coupled to a unimodal cavity. By exhaustively exploring a wide range of possible values for the hyperparameters using the Random Search Cross Validation method, we have found the optimal set of hyperparameters for three different mappings between the cumulants and the hot bath-induced coherence intervals. We have compared the different mappings and found that the validation accuracy of the classification models has $a_i(f_c)$ values greater than 81\%. Furthermore, we successfully developed a machine learning protocol to find the probability of predicting the hot bath-induced coherence intervals for a given set of cumulants, using all the classifiers. The classification results for each of the classifiers were consistent across the different cases examined, which highlights the robustness of the classifiers in predicting the coherence intervals for hot-bath-induced coherence. For a specified engine,  the trained model revealed that a photon flux larger (smaller) than the variance of photons during work done have a tendency to favour lower (higher) value of noise induced coherence. This prediction is independent of the values of higher order cumulants. These findings could potentially have important implications in the field of quantum thermodynamics, where the ability to predict the coherence intervals in a hot-bath-induced system could aid in the design and optimization of quantum thermal machines.

\begin{acknowledgments}
  MJS and HPG acknowledge the support from Science and Engineering Board, India for the start-up grant, SERB/SRG/2021/001088. HPG also acknowledges the support from University Grants Commission, New Delhi
for the startup research grant, UGC(BSR), Grant No. F.30-585/2021(BSR).
\end{acknowledgments}

\bibliography{rsc}

\end{document}